\documentclass[cameraready]{Interspeech}

\title{Is Word Error Rate Enough? Rethinking Privacy Evaluation in Speech with Entity-Aware Metrics }

\author[affiliation={1}, orcid=0009-0004-9291-2708, correspondingauthor]{Anjana}{Rajasekhar}
\author[affiliation={2,3}, orcid=0009-0000-6687-7559]{
Jule}{Pohlhausen}
\author[affiliation={1}]{Nayana Jacob}{Alappattu}
\author[affiliation={1},
orcid=0000-0003-2994-2336]{Anna}{Leschanowsky}

\address{
    $^1$ Fraunhofer Institute for Integrated Circuits (IIS), Erlangen, Germany \\
    $^2$ Institute of Hearing Technology and Audiology, Jade University of Applied Sciences, Oldenburg,
Germany \\
    $^3$ Dept. of Medical Physics and Acoustics, Carl von Ossietzky Universität Oldenburg, Germany
}

\email{anjana.rajasekhar@iis.fraunhofer.de}

\keywords{speech privacy, speech recognition, obfuscation, named entity recognition, attack models}

\usepackage{comment}
\usepackage{siunitx}
\usepackage{subcaption}
\usepackage{csquotes}
\usepackage{xcolor}

\begin{document}

\maketitle

\begin{abstract}
As the use of smart devices continues to increase, their potential to capture sensitive speech content raises growing privacy concerns. It is therefore critical to develop techniques that prevent information leakage while preserving the utility of the audio, and evaluation metrics that accurately quantify the level of privacy without overestimating it.
In this work, we evaluate the effectiveness of two obfuscation techniques in protecting speech content, with particular emphasis on named entities, by adapting entity-aware privacy metrics from the Natural Language Processing field to the speech privacy domain. Further, we investigate several attack scenarios and show that fine-tuning on entity-rich data improves attack performance for some entity categories but not others. Finally, we provide guidance on metric selection based on whether the obfuscation method preserves temporal alignment. 

\end{abstract}

\section{Introduction}
Over the years, smart devices have significantly improved human interaction with technology. These devices are capable of understanding their surroundings and detecting specific events in environments such as smart homes~\cite{hollosi2010voice}, elder care facilities~\cite{yang2025real}, and smart cities~\cite{zhang2023automatic}. In addition, many voice-driven interactive technologies operate with always-listening functionality~\cite{nautsch2019preserving}.
However, the audio collected by such systems may contain sensitive information such as named entities (e.g., locations, dates, organisations), as well as financial or medical information~\cite{williams2021revisiting}. With the increasing performance of Automatic Speech Recognition (ASR) systems, the risk of such sensitive information being extracted and misused has also grown. Hence, regulations like the European General Data Protection Regulation (GDPR) are in place to protect the content from misuse~\cite{nautsch2019gdpr}.

To mitigate the misuse of sensitive content and to abide by such regulations, several Privacy Enhancing Technologies (PETs) have been proposed over the years. These include techniques such as obfuscation, content masking, encryption, adversarial learning, and distributed learning~\cite{nautsch2019preserving,sato2026task,ahmed2020preech}. However, many of these approaches are computationally expensive or require additional annotations, making them less suitable for resource-constrained edge devices. For instance, content masking methods often rely on the precise identification and localisation of target words, which require additional processing steps such as forced alignment or keyword spotting~\cite{williams2023exploratory}. Similarly, adversarial or distributed learning typically involves significant computational overhead. In this work, we focus on lightweight obfuscation techniques that can be efficiently implemented on edge devices while preserving the utility of the audio, such as temporal and spectral smoothing~\cite{pohlhausen2026csl}, resampling~\cite{pohlhausen2025resample}, or sound shredding~\cite{kumar2015sound}. 

In previous studies, privacy achieved by such techniques has typically been evaluated using Word Error Rate (WER) obtained from ASR systems~\cite{kumar2015sound,liang2020characterizing}. However, WER assigns equal importance to all words, even though only a small subset of the speech content may be sensitive~\cite{sasindran2023h}. Consequently, WER-based evaluations may misrepresent privacy achieved through obfuscation, as errors on non-sensitive words can inflate the overall error rate while sensitive entities may remain recognisable. In this work, we focus on named entities (e.g., person names, locations, organisations, and dates) as a proxy for sensitive content. While there is no general agreement in the Natural Language Processing (NLP) community on which linguistic elements are sensitive, since they vary across contexts and communities~\cite{hershcovich2022challenges}, named entities often encode Personally Identifying Information (PII). While previous studies have used Named Entity Recognition (NER) to detect and remove entities in speech, these efforts have primarily focused on identifying sensitive content rather than privacy evaluation~\cite{cohn2019audio}. To the best of our knowledge, entity-aware evaluation metrics have not been applied to speech obfuscation. 

To address this gap, we make two contributions. First, we analyse how attacker knowledge affects entity recovery under lightweight obfuscation techniques and show that fine-tuning on entity-rich data disproportionately improves recognition of domain-specific entities. This motivates our second contribution, where we adapt entity-aware evaluation metrics from NLP to the speech privacy domain. 
Specifically, we adapt an alignment-based metric and an alignment-free metric, as different obfuscation methods may either preserve temporal alignment (e.g., smoothing) or disrupt it entirely (e.g., shredding). Using the Spoken Language Understanding Evaluation (SLUE) benchmark suite~\cite{shon2022slue}, a dataset with annotated named entities, we provide practical guidance on metric selection based on obfuscation characteristics in Section~\ref{sec:results_metrics}. Section~\ref{sec:experimental_setup} describes our experimental setup, including obfuscation methods, attack models, and proposed metrics. We then present and discuss results in Section~\ref{sec:results}.


 
\section{Experimental Setup}
\label{sec:experimental_setup}

The experiments were performed using one NVIDIA GeForce RTX 4090 GPU.
\subsection{Dataset}
The experiments were conducted using the Spoken Language Understanding Evaluation (SLUE) benchmark suite~\cite{shon2022slue}. The benchmark comprises subsets of the VoxCeleb~\cite{nagrani17_interspeech} and VoxPopuli~\cite{wang2021voxpopuli} corpora. We use the SLUE-VoxPopuli subset for fine-tuning and evaluation, as it was specifically designed for NER. SLUE-VoxPopuli provides speech recordings with their corresponding transcripts and manual annotations of the named entities contained in each utterance. Table~\ref{tab:slue_stats} gives the statistics for the fine-tune and test sets of the SLUE-VoxPopuli dataset.

\begin{table}[th]
\caption{Statistics of SLUE-VoxPopuli dataset.}
\label{tab:slue_stats}
\centering
\begin{tabular}{p{3cm} p{2cm} p{1.7cm}}
\toprule
Parameter & Fine-tune set & Test set \\
\midrule
Total recordings   & 5000  & 1842  \\
Nr. of speakers    & 139 & 275 \\
Total duration     & \SI{14.5}{\hour}  & \SI{5}{\hour}  \\
Mean duration      & 10.41 & 9.63   \\
\quad in s [Min, Max] & [0.62, 61.22]  & [0.58, 69.24]  \\[0.1ex]
Mean Nr. of entities  & 1.16  & 1.09  \\  
\quad per rec. [Min, Max] & [0, 21]  & [0, 11]  \\      
\bottomrule
\end{tabular}
\end{table}

\subsection{Obfuscation techniques}
We evaluate two lightweight, publicly available obfuscation techniques to assess how the protection of named entities compares to that of other words. The first is sound shredding, which segments the audio signal into fixed-length blocks, randomises their order, and concatenates the shuffled segments to reconstruct the signal~\cite{kumar2015sound}. Shredding has been proven to destroy speech content while preserving the acoustic features for Sound Event Classification (SEC)~\cite{jacobs2023speech}. Shredding distorts the temporal alignment with the transcript and is hence useful for evaluating the alignment-free metrics. In our experiments, we consider block lengths of \SIlist{0.5;1;2;5;10}{\second}. The chosen values span a broad range of shredding levels and allow us to investigate the sensitivity of privacy to block length. We consider ignorant attack models to evaluate this method as explained in section~\ref{sec:attack_model}. 

The second method is temporal smoothing with subsampling, which reduces the temporal resolution of the Power Spectral densities (PSD) before calculating Mel filterbank energies. Temporal smoothing with subsampling maintains temporal alignment and has proven effective for preserving the privacy of speech content in the context of conversation analysis~\cite{pohlhausen2026csl}. We evaluate smoothing time constants of $\tau = \SIlist{125;250;375;500}{\milli\second}$, corresponding to subsampling factors of 10, 20, 30, and 40. These values were shown to increase WER, with the WER on LibriSpeech test-clean exceeding 78~\% for $\tau \geq \SIlist{375}{\milli\second}$ when employing informed ASR models~\cite{pohlhausen2026csl}. We consider informed attack models to evaluate this method as explained in section~\ref{sec:attack_model}.

\subsection{Attack Models}
\label{sec:attack_model}
Considering different attack models is essential for evaluating speech obfuscation methods, as the level of attacker knowledge strongly influences the feasibility of recovering sensitive information. An ignorant attacker is unaware of the obfuscation method used, whereas an informed attacker knows the method and its exact parameter values~\cite{9053868}. An informed attack condition represents the strongest privacy concern associated with the obfuscation technique.
For sound shredding, we consider Whisper Base~\cite{radford2022whisper} and SpeechBrain-TransformerLM~\cite{speechbrain} as ignorant attack models, denoted as \textit{Whisper} and \textit{SpeechBrain-TF}, respectively, in the remainder of this work.
Both ASR systems use 80-dimensional log-Mel filterbank features computed using a 25~ms window and a 10~ms hop size. Since sound shredding breaks the temporal alignment between the speech signal and the corresponding transcript, training an ASR model on obfuscated data is not meaningful. For this type of obfuscation, a signal processing-based attack that attempts to reconstruct the original temporal ordering may be more appropriate. For example, a heuristic reconstruction algorithm could estimate the most likely ordering of the shuffled blocks by exploiting signal continuity or other acoustic cues, similar to approaches used in classical descrambling attacks in telephony~\cite{zhao2007puzzle}. As such attacks would make comparison with the fine-tuned ASR models used for temporal smoothing difficult, we restrict our analysis to ignorant attack models for shredding. 

For temporal smoothing with subsampling, we consider both ignorant and informed attack settings. To ensure comparison across obfuscation techniques, we again use \textit{Whisper} and \textit{SpeechBrain-TF} as ignorant attack models.
For the informed ASR model, we use a publicly available experimental framework\footnote{{https://github.com/ol-MEGA/ppca}}. We consider two informed attack variants that differ in their level of knowledge: (i) Obfuscation-aware attack model, with models trained on the obfuscated LibriSpeech train-clean-360 dataset~\cite{Panayotov2025librispeech} for each smoothing parameter over 90 epochs. This attack model knows the obfuscation method and its parameters, but is not adapted to the target domain, including limited knowledge of entities. (ii) Obfuscation- and domain-aware attack model, with models fine-tuned for an additional 30 epochs on the corresponding obfuscated SLUE fine-tuning set~\cite{shon2022slue}. This attack model has additional knowledge of the target domain, including exposure to named entities that appear in the SLUE corpus. This distinction allows us to analyse the effect of attacker knowledge on privacy evaluation, including the impact of entity knowledge, by comparing recognition performance on entities seen during training versus unseen entities.

\begin{table*}[t]
    \centering
    \caption{Total WER in~\% on SLUE test set~\cite{shon2022slue} for two obfuscation methods: 
    shredding (different block lengths) and temporal smoothing with subsampling (different $\tau$ values). 
    Informed attack models obtained by training SpeechBrain-TF with the obfuscated LibriSpeech 
    train-clean-360~\cite{Panayotov2025librispeech} set and corresponding $\tau$ values. 
    Additional reduction in WER was obtained by fine-tuning those models using the corresponding 
    obfuscated SLUE fine-tune set.}
    \label{tab:combined_results}
    \begin{tabular}{l l p{3.5cm} c ccccc}
        \toprule
        \textbf{Attack type} & \textbf{Model} & \textbf{Training set} & \textbf{Orig.} 
            & \multicolumn{5}{c}{\textbf{Obfuscated SLUE test set}} \\
        \midrule
        \multicolumn{9}{l}{\textit{Shredding}} \\
        \cmidrule(lr){5-9}
        & & & & \textbf{\SI{10}{\second}} & \textbf{\SI{5}{\second}} & \textbf{\SI{2}{\second}} 
            & \textbf{\SI{1}{\second}} & \textbf{\SI{0.5}{\second}} \\
        \cmidrule(lr){5-9}
        Ignorant & Whisper        & -- & 27.00 & 35.22 & 52.77 & 73.06 & 86.98 & 102.44 \\
        Ignorant & SpeechBrain-TF & -- & 24.53 & 35.12 & 54.58 & 82.00 & 97.50 & 113.20 \\
        \midrule
        \multicolumn{9}{l}{\textit{Temporal Smoothing with Subsampling}} \\
        \cmidrule(lr){5-8}
        & & & & \textbf{\SI{125}{\milli\second}} & \textbf{\SI{250}{\milli\second}} 
            & \textbf{\SI{375}{\milli\second}} & \textbf{\SI{500}{\milli\second}} & \\
        \cmidrule(lr){5-8}
        Ignorant & Whisper        & --                                          & 27.00 & 99.90 & 99.87 & 99.83 & 125.03 & \\
        Ignorant & SpeechBrain-TF & --                                          & 24.53 & 99.77 & 104.33 & 101.73 & 104.84 & \\
        Informed & SpeechBrain-TF & Obf. LS train-clean-360                     & --    & 60.37 & 106.91 & 123.40 & 126.41 & \\
        Informed & SpeechBrain-TF & Obf. LS train-clean-360 + FT on SLUE FT set & --    & 41.34 & 78.70 & 97.36 & 112.51 & \\
        \bottomrule
    \end{tabular}
    \par\smallskip
    {Obf: Obfuscated, LS: LibriSpeech, FT: Fine-tuning, Orig.: Original SLUE test set}
\end{table*}

\begin{figure*}[ht]
  \centering
  \begin{subfigure}[b]{0.48\textwidth}
    \centering
    \includegraphics[width=\textwidth]{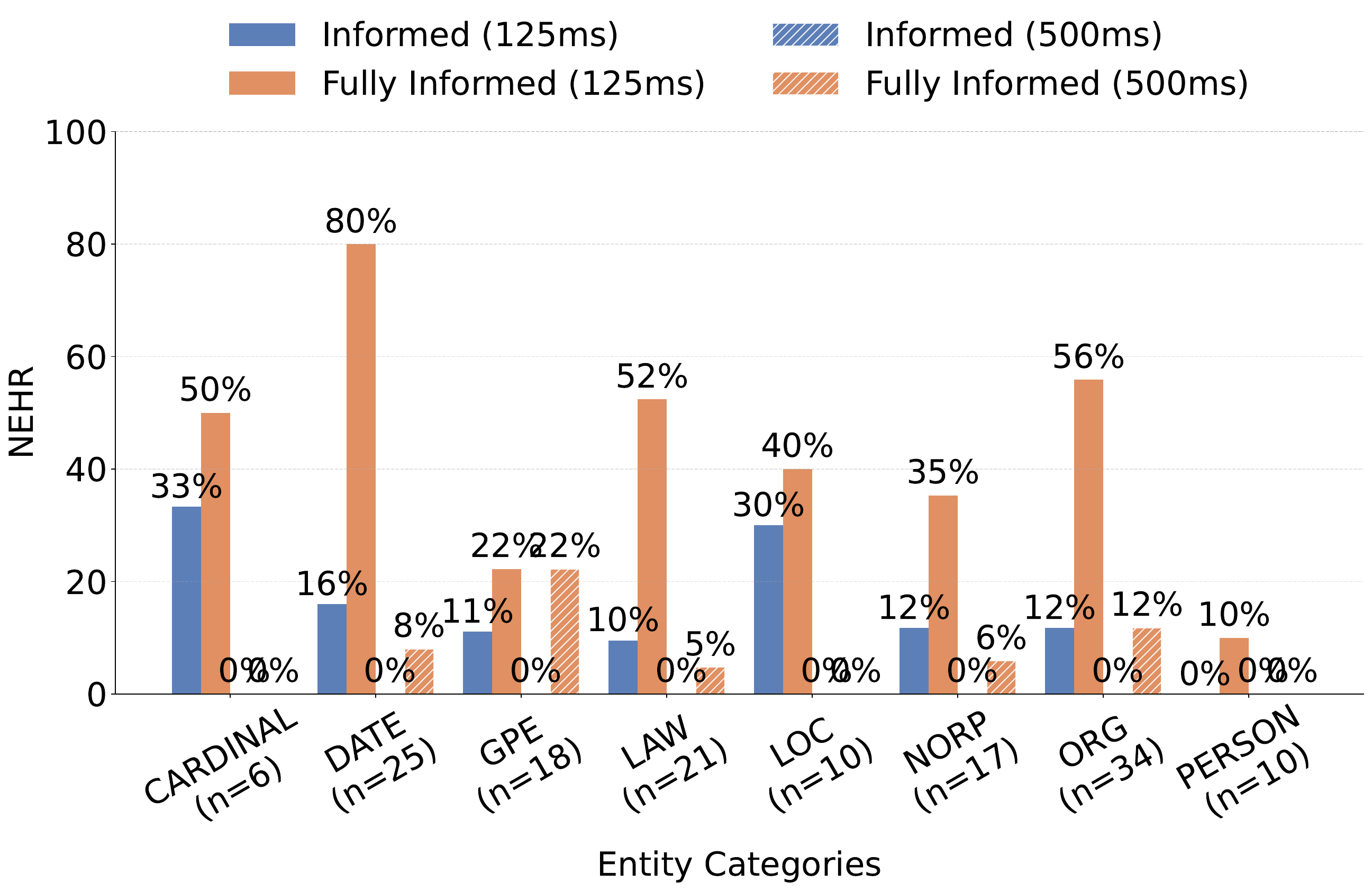}
    \caption{Percentage of test entities recognised by the informed attack model and the fully informed attack model for temporal smoothing with subsampling at $\tau = \SI{125}{\milli\second}$ and $\tau = \SI{500}{\milli\second}$. Entities that are present only in the SLUE fine-tune set are considered here.}
    \label{fig:entity_1_minus_nelr}
  \end{subfigure}
  \hfill
  \begin{subfigure}[b]{0.48\textwidth}
    \centering
    \includegraphics[width=\textwidth]{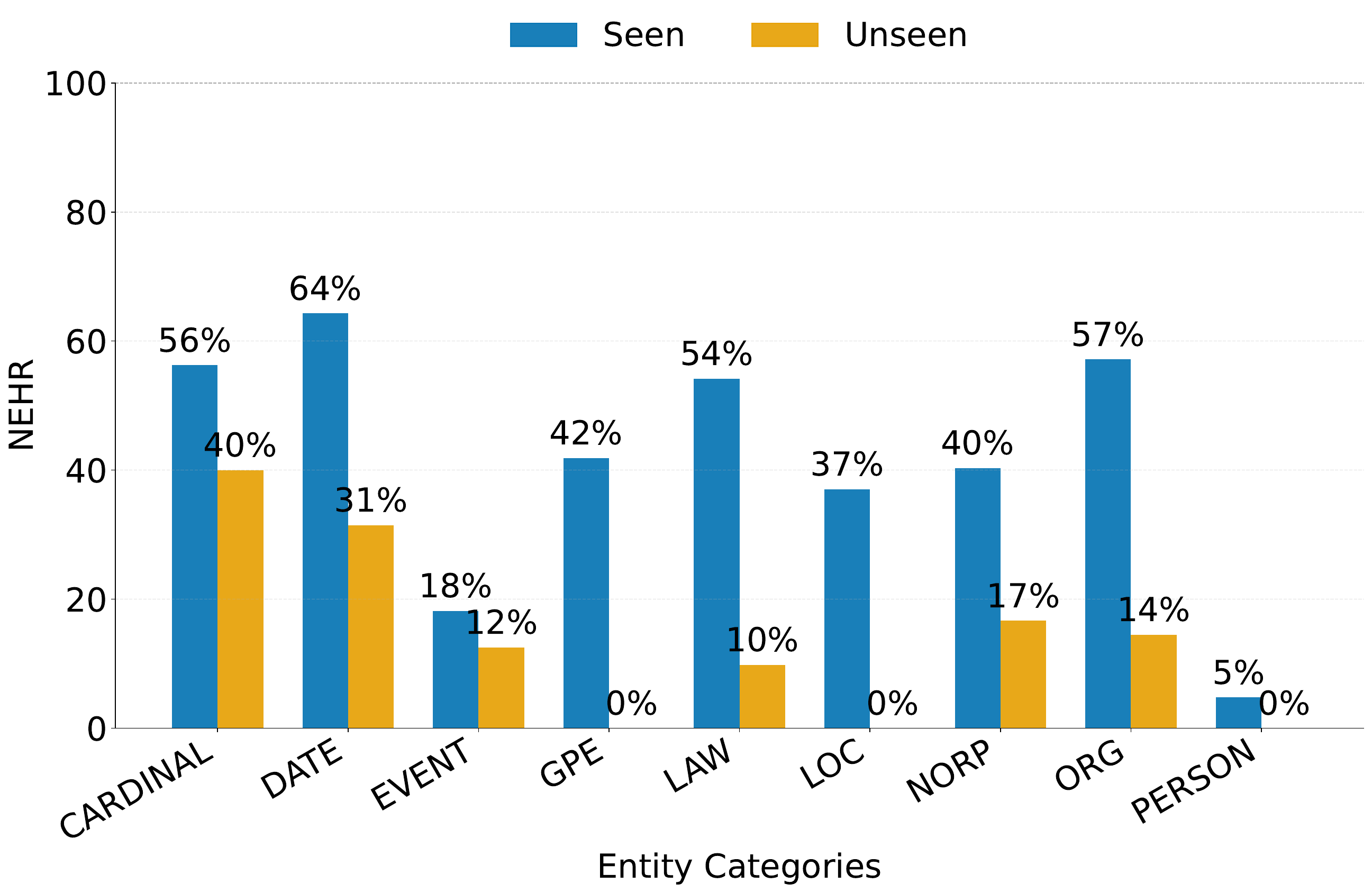}
    \caption{Percentage of seen and unseen test entities recognised by the fully informed attack model at $\tau = \SI{125}{\milli\second}$. Only categories with more than 10 seen entities are shown.}
    \label{fig:combined_entity_1_minus_nelr_125ms}
  \end{subfigure}
  \caption{NEHR by category for temporal smoothing with subsampling: (a) comparison between informed and fully informed attack models at two smoothing levels; (b) comparison between seen and unseen entities for the fully informed attack model.}
  \label{fig:entity_analysis}
\end{figure*}

\subsection{Entity-Aware Metrics}
\label{sec:metrics}



\subsubsection{Named Entity Error Rate (NEER)}
Standard WER treats all words equally, assigning the same penalty to errors on function words as to errors on named entities~\cite{favre2013automatic}. 
In spoken language understanding, Slot Error Rate (SER) addresses this by restricting error counting to slot values, where multi-word entities can be treated as single units~\cite{makhoul1999performance}. We adapt this principle to speech privacy evaluation by proposing the Named Entity Error Rate (NEER), which restricts error counting to reference tokens that belong to annotated named entity spans. 

To compute NEER, we first identify entity tokens in the reference transcript. Standard Levenshtein alignment is then performed on the complete reference and hypothesis token sequences to ensure reliable token correspondence~\cite{levenshtein1966binary}. Finally, substitutions and deletions are accumulated only at positions corresponding to reference entity tokens and are normalised by the total number of entity tokens:
\begin{equation}
\mathrm{NEER} = \frac{S_E + D_E}{N_E},
\label{eq:neer}
\end{equation}
where $S_E$ and $D_E$ denote the numbers of substitutions and deletions associated with reference entity tokens, respectively, and $N_E$ is the total number of entity tokens in the reference transcript. Unlike standard SER, we exclude insertions because they cannot be attributed to reference named entity positions and thus provide no information about whether the original entities were leaked or protected.

\subsubsection{1 - Named Entity Leakage Rate ($1 -$ NELR)}
While NEER may be well-suited for alignment-preserving obfuscation techniques such as temporal smoothing, it may overestimate privacy for methods that disrupt temporal structure, such as sound shredding. Here, entity tokens may still be recognisable but appear at unexpected positions in the hypothesis. To address this, we additionally propose $1 -$ NELR, which captures entity leakage independent from alignment. Unlike NEER, which relies on Levenshtein alignment and therefore restricts error counting to tokens along the alignment path, $1 -$ NELR does not use alignment. Instead, it performs a direct search over the ASR hypothesis to determine which unique reference entities appear in the output. That is, each named entity is counted only once, regardless of how often it appears in the reference sentence, and an entity is marked as leaked if it appears in the hypothesis sentence. $1 -$ NELR thus measures the proportion of unique named entities in the reference sentence that are absent from the ASR hypothesis sentence. A value of 0 indicates that all reference entities are leaked, whereas a value of 1 corresponds to complete protection, with none of the entities appearing in the hypothesis.
\begin{equation}
    \text{$1 -$ NELR} = 1 - \frac{
        \text{no. of unique reference entities leaked}
    }{
        \text{total no. of unique reference entities}
    }
\end{equation}

\subsubsection{Fuzzy matching based $1 -$ NELR}
To account for transcription errors associated with entities in the ASR output, we extend the $1 -$ NELR with fuzzy string matching~\cite{bosker2021using}. Exact string matching may fail to identify leaked entities when the entity is transcribed with minor spelling errors. We therefore consider an entity to be leaked if either the exact entity string or a sufficiently similar string occurs in the ASR hypothesis. Fuzzy matching is performed using the \texttt{rapidfuzz}\footnote{https://github.com/rapidfuzz/RapidFuzz} library. Specifically, we use the \texttt{fuzz.ratio} function, which computes a normalised character-level similarity score between 0 and 100 based on the edit similarity of the two strings as below:
\begin{equation}
\text{fuzz.ratio}(ref,hyp) = \frac{2 \cdot M}{|ref| + |hyp|} \times 100
\label{eq:fuzzratio}
\end{equation}
where $M$ is the number of matching characters between the reference ($ref$) word and the hypothesis ($hyp$) word found using the edit distance alignment. The resulting \texttt{fuzz.ratio} value represents the similarity between the two words on a scale from 0 to 100. For a given threshold $\theta$, an entity is considered leaked when its similarity score satisfies \texttt{fuzz.ratio} $\geq$ $\theta$. Thus, $\theta$ defines the minimum similarity required for the hypothesis word to be considered a match to the reference entity. As no single threshold value is universally established for this purpose, we evaluate leakage across a range of threshold values, considering $\theta \in \{10, 20, 30, \ldots, 100\}$.

\subsubsection{Named Entity Hit Rate (NEHR)}
Named Entity Hit Rate (NEHR) helps to analyse attack model strength rather than evaluating the performance of the obfuscation techniques. While $1 -$ NELR is computed per sentence, NEHR operates globally across the entire test set. Specifically, all unique entities in the test set are listed, and an entity is counted as ``hit'' if it is recognised at least once by the attack model. A high NEHR thus indicates that the attack model has sufficient entity recognition capability and can pose a privacy threat.
We define it as: 
\begin{equation}
    \text{NEHR} = \frac{
        \text{no. of unique test set entities recognised by attacker}
    }{
        \text{total no. of unique test set entities}
    }
\end{equation}

\section{Results and Discussions}
\label{sec:results}

\subsection{Attack Models}
\label{sec:results_attack}

\begin{figure*}[ht]
  \centering \includegraphics[width=1\textwidth]{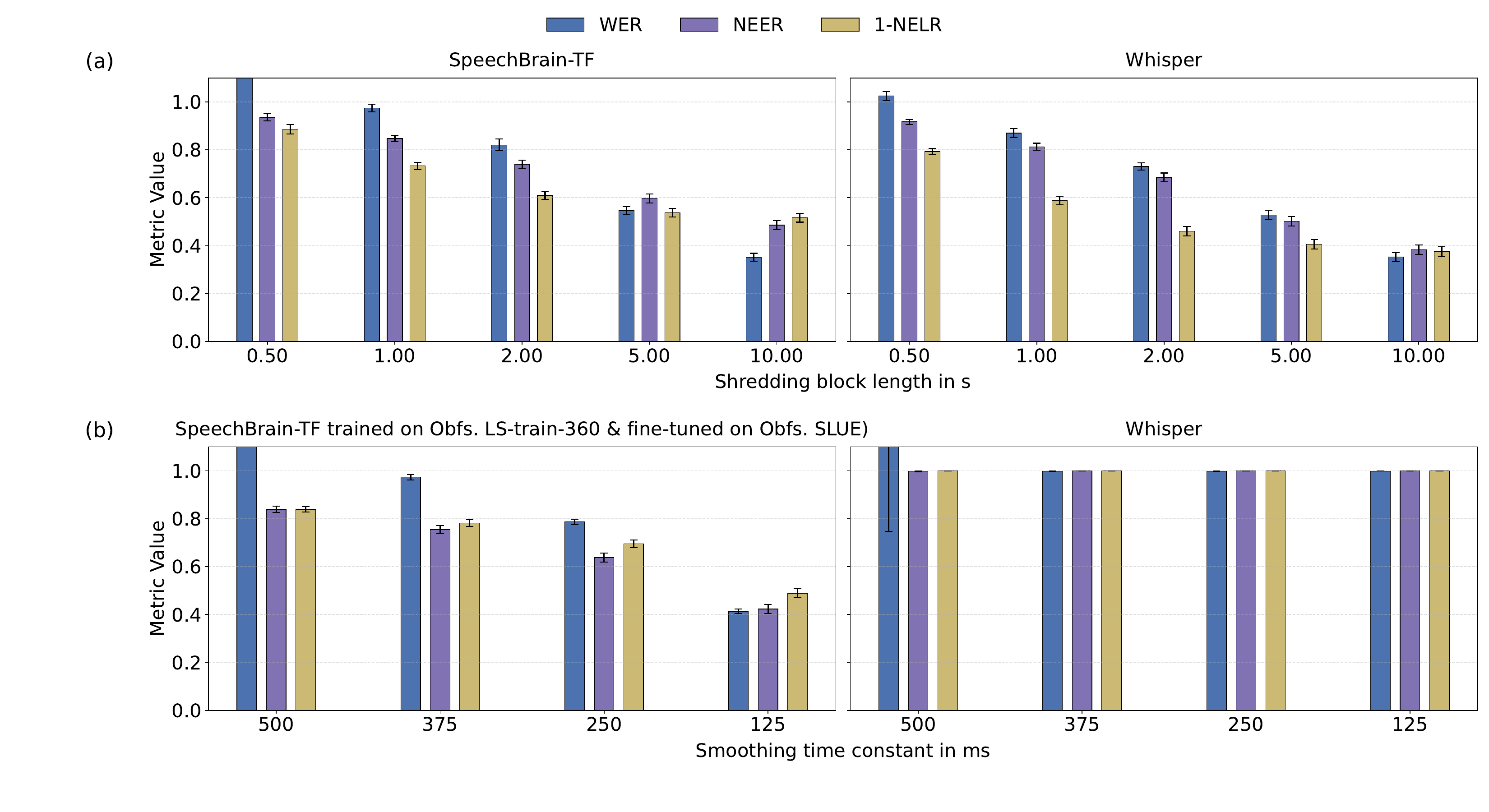}
  \caption{WER, NEER and $1 -$ NELR values for shredding and temporal smoothing with subsampling are reported for each parameter setting. For shredding, results are shown for both the pretrained SpeechBrain-TF and Whisper models. For temporal smoothing with subsampling, results are presented for the best-performing model alongside Whisper.}
  \label{fig:metrics}
\end{figure*}

Table~\ref{tab:combined_results} summarises the WER achieved by each attack type for different parameter values of shredding and temporal smoothing with subsampling, respectively. The performance of Whisper and SpeechBrain-TF on the original SLUE test set is consistent with the results reported in the SLUE benchmark~\cite{shon2022slue}, confirming that our baseline reproduces expected ASR performance. For shredding, the two ignorant attack models exhibit nearly identical performance, with the WER decreasing as the block length increases. Here, the ignorant attack establishes a baseline for evaluating potential signal-processing-based informed attack models.

For temporal smoothing with subsampling, ASR performance declines as the temporal smoothing constant increases. A clear performance gap is observed between the ignorant and informed attack models for low smoothing constants. At $\tau = \SI{125}{\milli\second}$, the SpeechBrain-TF model retrained on the obfuscated LibriSpeech train-clean-360 dataset achieves a WER that is nearly 40\% lower than that of the pretrained SpeechBrain-TF model. A further reduction in WER of approximately 20\% is obtained by fine-tuning the retrained model on the obfuscated SLUE dataset. To investigate the source of this improvement caused by fine-tuning, we separately evaluate the hit rate for entities and non-entities. The NEHR is extended to the non-entity span to analyse the non-entity recognition capability of the attacker models. A non-entity is considered a "hit" if it is recognised at least once by the attack model. From this point onward, we refer to the SpeechBrain-TF model retrained on the obfuscated LibriSpeech train-clean-360 dataset as the \enquote{informed attack model}, while the corresponding model further fine-tuned on the obfuscated SLUE fine-tuning set is referred to as the \enquote{fully informed attack model}, as it is additionally adapted to the target domain.

For temporal smoothing with subsampling at $\tau = \SI{125}{\milli\second}$, the hit rate for non-entities is 59\% for the informed attack model and 68\% for the fully informed attack model. In contrast, the hit rate for entities increases substantially, from 38\% for the informed attack model to 63\% for the fully informed attack model. These results indicate that the reduction in WER achieved through fine-tuning seems to be primarily due to the improved entity recognition. Hence, we calculate the NEHR to understand how strong the attack models are. Figure~\ref{fig:entity_1_minus_nelr} illustrates the NEHR across entity categories for smoothing constants of 125~ms and 500~ms. For both values, the fully informed attack model achieves equal or higher hit rates across all categories compared to the informed attack model, with more pronounced differences at the lower smoothing constant. While improvement can be observed across all entity categories, a clear recognition improvement is observed for categories like DATE, LAW and ORG.

To understand whether the improved hit rate is driven by exposure to new domain-specific entities, we separately evaluate entities present in both datasets, i.e., LibriSpeech train-clean-360 and SLUE fine-tuning set, and those unique to the SLUE fine-tuning set. 
For entities from the SLUE test set that are present in both training datasets, the recognition accuracy remains essentially unchanged, with the fully informed attack model achieving a hit rate of 62\% compared to 60\% for the informed attack model. Thus, fine-tuning does not improve the recognition of entities that were already observed during retraining on the obfuscated LibriSpeech dataset. In contrast, for entities appearing exclusively in the SLUE fine-tuning set, the recognition accuracy increases substantially, from 13\% for the informed attack model to 46\% for the fully informed attack model. This indicates that the overall WER reduction achieved through fine-tuning is primarily driven by improved recognition of entities that are unique to the target domain and were not encountered during retraining on the obfuscated LibriSpeech dataset.

\begin{table}[!b]
\caption{Statistics of test-set named entities with respect to the training data used by the fully informed model for temporal smoothing with subsampling. Seen entities appear in the LibriSpeech train-clean-360 dataset or the SLUE fine-tuning set or both. Unseen entities appear only in the test set.}
\label{tab:entity_stats}
\centering
\begin{tabular}{l p{1cm} p{1cm} p{1.3cm} p{1.5cm}}
\toprule
Entity Categories & Test Entities & Seen Entities (\%) & Unseen Entities (\%)\\
\midrule
CARDINAL    & 84  & 76.19 & 23.81\\
DATE        & 119 & 70.58 & 29.42\\
EVENT       & 19  & 57.89 & 42.11\\
FAC         & 4   & 25.00 & 75.00\\
GPE         & 120 & 71.66 & 28.34\\
LANGUAGE    & 7   & 100.00 & 0.00\\
LAW         & 75  & 32.00 & 68.00\\
LOC         & 34  & 79.41 & 20.59\\
MONEY       & 6   & 33.33 & 66.67\\
NORP        & 91  & 73.62 & 26.38\\
ORDINAL     & 10  & 100.00  & 0.00\\
ORG         & 125 & 39.20 & 60.80\\
PERCENT     & 2   & 100.00  & 0.00\\
PERSON      & 59  & 35.59  & 64.41\\
PRODUCT     & 6   & 16.67 & 83.33\\
QUANTITY    & 8   & 50.00  & 50.00\\
TIME        & 6   & 100.00  & 0.00\\
WORK\_OF\_ART & 3 & 0.00 & 100.00\\
\midrule
\textbf{Total} & \textbf{778} & \textbf{59.90} & \textbf{40.10}\\
\bottomrule
\end{tabular}
\end{table}

We next investigate how prior exposure to entities during training affects recognition performance by considering the fully informed attack model, which is the strongest attack model in the current setup. Specifically, we compare the NEHR for entities that were seen during training, i.e., present in LibriSpeech train-clean-360 and/or SLUE fine-tuning dataset, with unseen entities that were not observed during training. Table~\ref{tab:entity_stats} summarises the proportion of seen and unseen test entities per category. Overall, approximately 60\% of the test entities were seen during training. The NEHR for both groups are illustrated by Figure~\ref{fig:combined_entity_1_minus_nelr_125ms} for the fully informed attack model at $\tau = \SI{125}{\milli\second}$. Categories without seen or unseen entities, or with fewer than 10 entities, are omitted to ensure meaningful comparison. Results for other smoothing constants exhibit the same trend and are omitted due to space constraints.

\begin{figure*}[ht]
  \centering \includegraphics[width=1\textwidth]{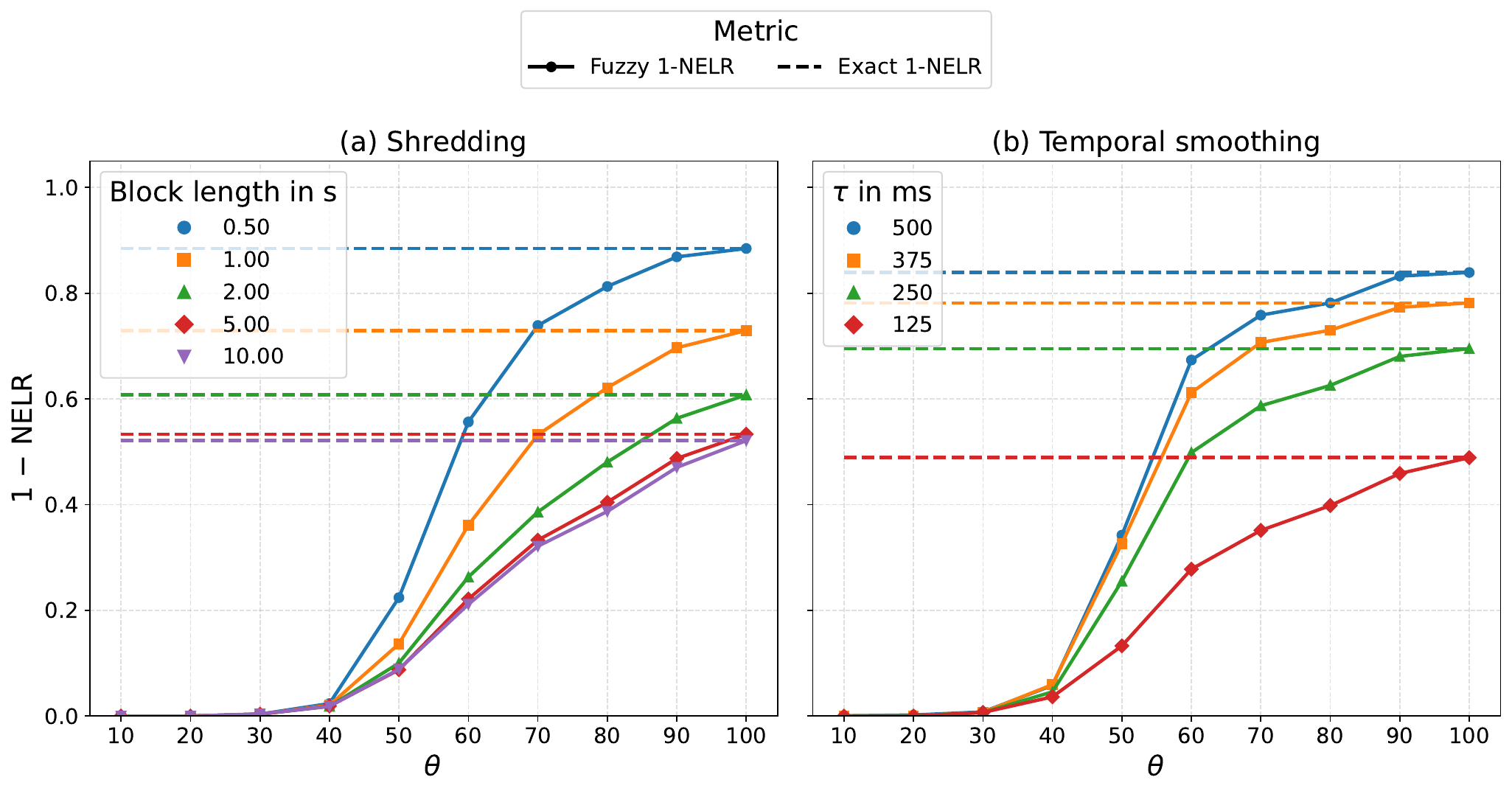}
  \caption{Effect of fuzzy matching threshold ($\theta$) and technique-specific parameters on $1 -$ NELR. (a) Shredding with varying block lengths for pretrained SpeechBrain-TF. (b) Temporal smoothing with varying smoothing parameter $\tau$ for the best-performing model. Solid lines with circular markers show $1 -$ NELR under fuzzy matching across different thresholds, while dashed horizontal lines show the corresponding $1 -$ NELR under exact matching.}
  \label{fig:fuzzy}
\end{figure*}

Across all entity categories, unseen entities display on average 28\% lower hit rate than seen entities. This demonstrates that prior exposure substantially improves recognition, whereas entities not encountered during training remain considerably more difficult for the attack model to recognise. However, this difference is less pronounced for categories of EVENT and PERSON, which exhibit low recognition accuracy even for seen entities. Notably, for the PERSON category, only approximately 5\% of entities that were seen during training get recognised. This result suggests that generic attack models struggle with certain categories even after domain adaptation. Category-specific attack models may therefore provide a more realistic assessment of the privacy risks associated with an individual category, particularly for sensitive categories such as PERSON. 

\subsection{Evaluation Metrics}
\label{sec:results_metrics}

We now evaluate the impact of the proposed entity-aware metrics on performance. Figure~\ref{fig:metrics} depicts the WER and entity-aware metrics values for shredding and temporal smoothing with subsampling. WER is consistently higher than NEER and $1 -$ NELR except for the 5 and \SI{10}{\second} setting in shredding and \SI{125}{\milli\second} setting in temporal smoothing. We suspect that, at these settings, there are fewer insertions and non-entity related transcription errors, which lowers WER, but not entity-aware metric values. 
This has implications for privacy assessment. In highly obfuscated cases, WER is inflated and shows higher values than entity-aware metrics, possibly overestimating privacy protection. In less obfuscated cases, the entity metrics can exceed WER, revealing entity-specific recognition errors. Here, WER may underestimate privacy protection with respect to sensitive information.


Comparing the two entity-aware metrics, we find that both yield comparable results for temporal smoothing, an alignment-preserving obfuscation technique. However, for shredding, which disrupts temporal alignment, $1 -$ NELR indicates higher leakage than NEER. This difference is more pronounced for the Whisper model. This suggests that the alignment-based NEER metric fails to capture some entity leakage when temporal structure is broken, potentially overestimating privacy protection for such methods. We therefore recommend using the alignment-free $1 -$ NELR metric for evaluating obfuscation methods that disrupt alignment, while either metric seems appropriate for alignment-preserving methods.

Figure~\ref{fig:fuzzy} further investigates the effect of fuzzy matching on $1 -$ NELR by depicting the $1 -$ NELR obtained for different fuzzy matching threshold values ($\theta$) for shredding and temporal smoothing. Fuzzy matching is used to identify entities that may have been modified or obfuscated but remain sufficiently similar to their original forms, thereby capturing leakage that would not be detected by exact matching. Entities are considered to be leaked when \texttt{fuzz.ratio} is greater than or equal to the respective $\theta$ values. For both techniques, $1 -$ NELR is near zero for $\theta$ below 40, rises sharply between 40 and 60, and converges toward the exact-match value (dashed lines) by 90 and 100. Nearly all entities are counted as leaked regardless of the obfuscation applied for $\theta$ below 40. This behaviour can be attributed to the character-level similarity measure used by \texttt{fuzz.ratio}. At low thresholds, short strings can receive relatively high similarity scores from a small amount of character-level overlap, even when the strings are not semantically related. As $\theta$ increases beyond 40, the matching criterion becomes stricter, allowing the metric to better distinguish between genuinely similar strings and coincidental character overlap. Consequently, $1-$ NELR increases sharply between $\theta$ values of 40 and 60. At high thresholds of 90 and 100, fuzzy matching approaches the behaviour of exact matching, causing the metric to converge toward the exact-match value. A fuzzy-matching threshold in the range of 60 to 80 appears to provide a reasonable trade-off between coincidental matching and exact matching. Whether a threshold within the proposed range increases privacy risks or makes it easier for an attacker to guess the correct word remains an open question that may require further investigation through human studies.

Our evaluation is restricted to a single alignment-preserving and a single alignment-free obfuscation method. While this enables a detailed analysis of their privacy properties, the findings may not generalise to other techniques that employ different transformation strategies. 
Future work should therefore investigate whether the observed results, in terms of both evaluation metrics and attack model behaviour, hold across a broader range of obfuscation methods. The evaluation is also constrained by the limited availability of speech datasets with entity-level annotations. A key limitation of the SLUE dataset is its narrow domain focus on European Parliament recordings. Having large-scale, diverse entity-annotated speech datasets would improve evaluation and enhance the generalisability of the results.

\section{Conclusion}
\label{sec:conclusion}
We evaluated two speech obfuscation techniques under different attack models and observed that fine-tuning an ASR model with an entity-rich dataset considerably improves the performance of the informed attack model, highlighting the importance of entity-aware training data in reconstruction attacks. The proposed entity-aware metrics provide additional insight into entity-level privacy protection, as WER may over- or underestimate privacy by including both entity and non-entity errors. Furthermore, our results show that $1 -$ NELR is particularly useful for obfuscation methods that disrupt alignment, as it captures all leaked entities, in contrast to NEER, which relies on alignment and may therefore miss certain leakage cases. In addition, fuzzy matching provides an alternative to exact matching by accounting for minor differences between entities, thereby enabling a more comprehensive assessment of leakage.

\section{Acknowledgements}
The authors thank Viktor Hangya for his valuable insights into the privacy metrics used in this work, and Meinard Müller for his helpful discussions and support throughout the work. 

\section{Generative AI Use Disclosure}
During manuscript preparation, the authors used AI-assisted writing tools to improve language and readability. After using these tools, the authors carefully reviewed and edited the content as necessary and take full responsibility for the content of the publication.

\bibliographystyle{IEEEtran}
\bibliography{mybib}

@inproceedings{Panayotov2025librispeech,
  author    = {Panayotov, Vassil and Chen, Guoguo and Povey, Daniel and Khudanpur, Sanjeev},
  booktitle = {Proc. {IEEE} Int. Conf. on Acoust., Speech and Signal Process. ({ICASSP})},
  title     = {Librispeech: An {ASR} corpus based on public domain audio books},
  year      = {2015},
  pages     = {5206-5210},
  doi       = {10.1109/ICASSP.2015.7178964}
}

@inproceedings{pohlhausen2025resample,
	title = {{Revisiting the Privacy of Low-Frequency Speech Signals: Exploring Resampling Methods, Evaluation Scenarios, and Speaker Characteristics}},
    author = {Pohlhausen, Jule and Bitzer, J{\"o}rg},
    booktitle = {{5th Symposium on Security and Privacy in Speech Communication}},
	publisher = {{ISCA}},
	pages = {85--89},
	doi = {10.21437/SPSC.2025-13},
  	year = {2025},
}

@misc{speechbrain,
  title={{SpeechBrain}: A General-Purpose Speech Toolkit},
  author={Mirco Ravanelli and Titouan Parcollet and Peter Plantinga and Aku Rouhe and Samuele Cornell and Loren Lugosch and Cem Subakan and Nauman Dawalatabad and Abdelwahab Heba and Jianyuan Zhong and Ju-Chieh Chou and Sung-Lin Yeh and Szu-Wei Fu and Chien-Feng Liao and Elena Rastorgueva and François Grondin and William Aris and Hwidong Na and Yan Gao and Renato De Mori and Yoshua Bengio},
  year={2021},
  eprint={2106.04624},
  archivePrefix={arXiv},
  primaryClass={eess.AS},
  note={arXiv:2106.04624}
}

@InProceedings{radford2022whisper,
  title = 	 {Robust Speech Recognition via Large-Scale Weak Supervision},
  author =       {Radford, Alec and Kim, Jong Wook and Xu, Tao and Brockman, Greg and Mcleavey, Christine and Sutskever, Ilya},
  booktitle = 	 {Proceedings of the 40th International Conference on Machine Learning},
  pages = 	 {28492--28518},
  year = 	 {2023},
  editor = 	 {Krause, Andreas and Brunskill, Emma and Cho, Kyunghyun and Engelhardt, Barbara and Sabato, Sivan and Scarlett, Jonathan},
  volume = 	 {202},
  series = 	 {Proceedings of Machine Learning Research},
  month = 	 {23--29 Jul},
  publisher =    {PMLR}
}

@inproceedings{williams2021revisiting,
  title     = {{Revisiting Speech Content Privacy}},
  author    = {Jennifer Williams and Junichi Yamagishi and Noe, Paul-Gauthier and Cassia Valentini-Botinhao and Jean-François Bonastre},
  year      = {2021},
  booktitle = {{2021 ISCA Symposium on Security and Privacy in Speech Communication}},
  pages     = {42--46},
  doi       = {10.21437/SPSC.2021-9},
}

@inproceedings{makhoul1999performance,
  title={Performance measures for information extraction},
  author={Makhoul, John and Kubala, Francis and Schwartz, Richard and Weischedel, Ralph and others},
  booktitle={Proceedings of DARPA broadcast news workshop},
  volume={249},
  pages={252},
  year={1999},
  organization={Herndon, VA}
}

@article{pohlhausen2026csl,
  title = {Towards privacy-preserving conversation analysis in everyday life: Exploring the privacy-utility trade-off},
journal = {Computer Speech \& Language},
volume = {95},
pages = {101823},
year = {2026},
issn = {0885-2308},
doi = {https://doi.org/10.1016/j.csl.2025.101823},
author = {Jule Pohlhausen and Francesco Nespoli and Jörg Bitzer}
}

@inproceedings{sato2026task,
  title={Task-Oriented Sound Privacy Preservation for Sound Event Detection Via End-to-End Adversarial Multi-Task Learning},
  author={Sato, Nao and Yasuda, Masahiro and Saito, Shoichiro},
  booktitle={ICASSP 2026-2026 IEEE International Conference on Acoustics, Speech and Signal Processing (ICASSP)},
  pages={14532--14536},
  year={2026},
  organization={IEEE}
}

@inproceedings{ahmed2020preech,
  title={Preech: A system for $\{$Privacy-Preserving$\}$ speech transcription},
  author={Ahmed, Shimaa and Chowdhury, Amrita Roy and Fawaz, Kassem and Ramanathan, Parmesh},
  booktitle={29th USENIX Security Symposium (USENIX Security 20)},
  pages={2703--2720},
  year={2020}
}

@inproceedings{nautsch2019gdpr,
  title     = {{The GDPR \& Speech Data: Reflections of Legal and Technology Communities, First Steps Towards a Common Understanding}},
  author    = {Andreas Nautsch and Catherine Jasserand and Els Kindt and Massimiliano Todisco and Isabel Trancoso and Nicholas Evans},
  year      = {2019},
  booktitle = {{Interspeech 2019}},
  pages     = {3695--3699},
  doi       = {10.21437/Interspeech.2019-2647},
  issn      = {2958-1796},
}

@inproceedings{shon2022slue,
  title={Slue: New benchmark tasks for spoken language understanding evaluation on natural speech},
  author={Shon, Suwon and Pasad, Ankita and Wu, Felix and Brusco, Pablo and Artzi, Yoav and Livescu, Karen and Han, Kyu J},
  booktitle={ICASSP 2022-2022 IEEE International Conference on Acoustics, Speech and Signal Processing (ICASSP)},
  pages={7927--7931},
  year={2022},
  organization={IEEE}
}

@inproceedings{hollosi2010voice,
  title={Voice activity detection driven acoustic event classification for monitoring in smart homes},
  author={Hollosi, Danilo and Schr{\"o}der, Jens and Goetze, Stefan and Appell, Jens-E},
  booktitle={3rd International Symposium on Applied Sciences in Biomedical and Communication Technologies (ISABEL)},
  pages={1--5},
  year={2010},
  organization={IEEE}
}

@article{yang2025real,
  title={Real-Time Acoustic Scene Recognition for Elderly Daily Routines Using Edge-Based Deep Learning},
  author={Yang, Hongyu and Dong, Rou and Guo, Rong and Che, Yonglin and Xie, Xiaolong and Yang, Jianke and Zhang, Jiajin},
  journal={Sensors},
  volume={25},
  number={6},
  pages={1746},
  year={2025},
  publisher={MDPI}
}

@article{zhang2023automatic,
  title={An Automatic Classification System for Environmental Sound in Smart Cities},
  author={Zhang, Dongping and Zhong, Ziyin and Xia, Yuejian and Wang, Zhutao and Xiong, Wenbo},
  journal={Sensors},
  volume={23},
  number={15},
  pages={6823},
  year={2023},
  publisher={MDPI}
}

@article{nautsch2019preserving,
  title={Preserving privacy in speaker and speech characterisation},
  author={Nautsch, Andreas and Jim{\'e}nez, Abelino and Treiber, Amos and Kolberg, Jascha and Jasserand, Catherine and Kindt, Els and Delgado, H{\'e}ctor and Todisco, Massimiliano and Hmani, Mohamed Amine and Mtibaa, Aymen and others},
  journal={Computer Speech \& Language},
  volume={58},
  pages={441--480},
  year={2019},
  publisher={Elsevier}
}

@INPROCEEDINGS{9053868,
  author={Lal Srivastava, Brij Mohan and Vauquier, Nathalie and Sahidullah, Md and Bellet, Aurélien and Tommasi, Marc and Vincent, Emmanuel},
  booktitle={ICASSP 2020 - 2020 IEEE International Conference on Acoustics, Speech and Signal Processing (ICASSP)}, 
  title={Evaluating Voice Conversion-Based Privacy Protection against Informed Attackers}, 
  year={2020},
  volume={},
  number={},
  pages={2802-2806},
  doi={10.1109/ICASSP40776.2020.9053868}}

@inproceedings{williams2023exploratory,
author={Williams, Jennifer and Pizzi, Karla and Noe, Paul-Gauthier and Das, Sneha},
  booktitle={Speech Communication; 15th ITG Conference}, 
  title={Exploratory Evaluation of Speech Content Masking}, 
  year={2023},

  volume={},

  number={},

  pages={215-219},

  doi={10.30420/456164042}}

@inproceedings{liang2020characterizing,
  title={Characterising the effect of audio degradation on privacy perception and inference performance in audio-based human activity recognition},
  author={Liang, Dawei and Song, Wenting and Thomaz, Edison},
  booktitle={22nd International Conference on Human-Computer Interaction with Mobile Devices and Services},
  pages={1--10},
  year={2020}
}

@inproceedings{sasindran2023h,
  title={H eval: A new hybrid evaluation metric for automatic speech recognition tasks},
  author={Sasindran, Zitha and Yelchuri, Harsha and Prabhakar, TV and Rao, Supreeth},
  booktitle={2023 IEEE Automatic Speech Recognition and Understanding Workshop (ASRU)},
  pages={1--7},
  year={2023},
  organization={IEEE}
}

@inproceedings{hershcovich2022challenges,
  title = "Challenges and Strategies in Cross-Cultural {NLP}",
    author = "Hershcovich, Daniel  and
      Frank, Stella  and
      Lent, Heather  and
      de Lhoneux, Miryam  and
      Abdou, Mostafa  and
      Brandl, Stephanie  and
      Bugliarello, Emanuele  and
      Cabello Piqueras, Laura  and
      Chalkidis, Ilias  and
      Cui, Ruixiang  and
      Fierro, Constanza  and
      Margatina, Katerina  and
      Rust, Phillip  and
      S{\o}gaard, Anders",
    editor = "Muresan, Smaranda  and
      Nakov, Preslav  and
      Villavicencio, Aline",
    booktitle = "Proceedings of the 60th Annual Meeting of the Association for Computational Linguistics (Volume 1: Long Papers)",
    month = may,
    year = "2022",
    address = "Dublin, Ireland",
    publisher = "Association for Computational Linguistics",
    doi = "10.18653/v1/2022.acl-long.482",
    pages = "6997--7013"
}

@inproceedings{kumar2015sound,
  author = {Kumar, Sumeet and Nguyen, Le T. and Zeng, Ming and Liu, Kate and Zhang, Joy},
title = {Sound Shredding: Privacy Preserved Audio Sensing},
year = {2015},
isbn = {9781450333917},
publisher = {Association for Computing Machinery},
address = {New York, NY, USA},
doi = {10.1145/2699343.2699366},
booktitle = {Proceedings of the 16th International Workshop on Mobile Computing Systems and Applications},
pages = {135–140},
numpages = {6},
location = {Santa Fe, New Mexico, USA},
series = {HotMobile '15}
}

@inproceedings{favre2013automatic,
  title={Automatic human utility evaluation of ASR systems: Does WER really predict performance?},
  author={Favre, Benoit and Cheung, Kyla and Kazemian, Siavash and Lee, Adam and Liu, Yang and Munteanu, Cosmin and Nenkova, Ani and Ochei, Dennis and Penn, Gerald and Tratz, Stephen and others},
  booktitle={INTERSPEECH},
  volume={13},
  pages={3463--3467},
  year={2013}
}

@inproceedings{cohn2019audio,
  title={Audio de-identification-a new entity recognition task},
  author={Cohn, Ido and Laish, Itay and Beryozkin, Genady and Li, Gang and Shafran, Izhak and Szpektor, Idan and Hartman, Tzvika and Hassidim, Avinatan and Matias, Yossi},
  booktitle={Proceedings of the 2019 Conference of the North American Chapter of the Association for Computational Linguistics: Human Language Technologies, Volume 2 (Industry Papers)},
  pages={197--204},
  year={2019}
}

@inproceedings{wang2021voxpopuli,
  title={VoxPopuli: A large-scale multilingual speech corpus for representation learning, semi-supervised learning and interpretation},
  author={Wang, Changhan and Riviere, Morgane and Lee, Ann and Wu, Anne and Talnikar, Chaitanya and Haziza, Daniel and Williamson, Mary and Pino, Juan and Dupoux, Emmanuel},
  booktitle={Proceedings of the 59th Annual Meeting of the Association for Computational Linguistics and the 11th International Joint Conference on Natural Language Processing (Volume 1: Long Papers)},
  pages={993--1003},
  year={2021}
}

@inproceedings{nagrani17_interspeech,
  title     = {{VoxCeleb: A Large-Scale Speaker Identification Dataset}},
  author    = {Arsha Nagrani and Joon Son Chung and Andrew Zisserman},
  year      = {2017},
  booktitle = {{Interspeech}},
  pages     = {2616--2620},
  doi       = {10.21437/Interspeech.2017-950},
  issn      = {2958-1796},
}

@inproceedings{jacobs2023speech,
  title={Speech Obfuscation in Mel Spectra that Allows for Centralised Annotation and Classification of Sound Events},
  author={Jacobs, Michiel and Vuegen, Lode and Khan, Suraj and Karsmakers, Peter and Make, Flanders},
  booktitle={Proceedings of the 8th Workshop on Detection and Classification of Acoustic Scenes and Events (DCASE 2023)},
  year={2023}
}

@inproceedings{zhao2007puzzle,
  title={A puzzle solver and its application in speech descrambling},
  author={Zhao, Yu-Xiang and Su, Mu-Chun and Chou, Zhong-Lie and Lee, Jonathan},
  booktitle={WSEAS international conference on computer engineering and applications},
  pages={171--176},
  year={2007}
}

@article{bosker2021using,
  title={Using fuzzy string matching for automated assessment of listener transcripts in speech intelligibility studies},
  author={Bosker, Hans Rutger},
  journal={Behavior Research Methods},
  volume={53},
  number={5},
  pages={1945--1953},
  year={2021},
  publisher={Springer}
}

@inproceedings{levenshtein1966binary,
  title={Binary codes capable of correcting deletions, insertions, and reversals},
  author={Levenshtein, Vladimir I and others},
  booktitle={Soviet physics doklady},
  volume={10},
  number={8},
  pages={707--710},
  year={1966},
  organization={Soviet Union}
}

\end{document}